\documentclass[showpacs,amssymb,amsmath,aps,prl,reprint,superscriptaddress]{revtex4-1}
\usepackage{graphicx}
\usepackage{bm}
\usepackage{dsfont}

\newcommand\I{\texttt{i}}

\newcommand\E{\mathrm{e}}

\newcommand\sech{\mathop{\rm sech}\nolimits}

\renewcommand\Im{\mathop{\rm Im}\nolimits}

\begin{document}
\title{The effect of the fifth-order nonlinearity on the existence of bright solitons below the modulation instability threshold}
\author{I.\,S.~Gandzha}
\author{Y\lowercase{u}.\,V.~Sedletsky}
\affiliation{Institute of Physics, Nat.~Acad.~of~Sci.~of Ukraine, Prosp.~Nauky 46, Kyiv 03028, Ukraine}
\email{gandzha@iop.kiev.ua, sedlets@iop.kiev.ua}

\date{\today}
\begin{abstract}
We analyze three different high-order nonlinear Schr\"{o}dinger
equation (HONLSE) models that have been used in the literature to
describe the evolution of slowly modulated gravity waves on the
surface of ideal finite-depth fluid. We demonstrate that the inclusion of the fifth-order nonlinear term to the HONLSE model introduces
only a small correction to the amplitude of the bright HONLSE soliton
solutions obtained without this term. Such soliton slutions behave as quasi-solitons in this more general case.
\end{abstract}
\maketitle

There are three different high-order nonlinear Schr\"{o}dinger
equation (HONLSE) models that have been used in the literature to
describe the evolution of slowly modulated gravity waves on the
surface of finite-depth irrotational, inviscid, and incompressible
fluid with flat bottom:
\begin{multline}\label{eq:Sedletsky}
u_\tau = - a_1 u_\chi - \I a_2 u_{\chi\chi}+\I a_{0,\,0,\,0}|u|^2 u \\
+ \Bigl(a_3 u_{\chi\chi\chi} - a_{1,\,0,\,0}u_{\chi}|u|^{2} - a_{0,\,0,\,1}u^{2}u^{*}_{\chi}\Bigr),
\end{multline}
\begin{multline}\label{eq:Slunyaev_full}
u_\tau = - a_1 u_\chi - \I a_2 u_{\chi\chi}+\I a_{0,\,0,\,0}|u|^2 u \\
+ \Bigl(a_3 u_{\chi\chi\chi} - \widetilde{a}_{1,\,0,\,0}u_{\chi}|u|^{2} - \widetilde{a}_{0,\,0,\,1}u^{2}u^{*}_{\chi}\Bigr)\\
+ \I \widetilde{a}_{0,\,0,\,0,\,0,\,0}|u|^4 u +
\I\Bigl(\widetilde{a}_4 u_{\chi\chi\chi\chi} - \widetilde{a}_{2,\,0,\,0}u_{\chi\chi}|u|^2 \\
- \widetilde{a}_{1,\,1,\,0}u_{\chi}^2 u^* - \widetilde{a}_{1,\,0,\,1}|u_\chi|^2 u - \widetilde{a}_{0,\,0,\,2}u^2 u_{\chi\chi}^*\Bigr),
\end{multline}
\begin{multline}\label{eq:Slunyaev_truncated}
u_\tau = - a_1 u_\chi - \I a_2 u_{\chi\chi}+\I a_{0,\,0,\,0}|u|^2 u \\
+ \Bigl(- \widetilde{a}_{1,\,0,\,0}u_{\chi}|u|^{2} - \widetilde{a}_{0,\,0,\,1}u^{2}u^{*}_{\chi}\Bigr)\\
+\I \widetilde{a}_{0,\,0,\,0,\,0,\,0}|u|^4 u,
\end{multline}
Here $u(\chi,\,\tau)$ is the first-harmonic envelope of the wave
profile, $\chi$ is the horizontal axis directed along the wave
propagation, and $\tau$ is time. The notation of equation coefficients
was selected such that the number of indices at coefficients
$a_{n}\ldots$ corresponds to the order of nonlinearity in the
corresponding term. The index values correspond to the orders of
derivatives present in that term.

Equation (\ref{eq:Sedletsky}) was
originally derived by Sedletsky \cite{SedletskyJETP2003} (2003) and
then was converted to dimensionless form by Gandzha et al.
\cite{UJP_2014} (2014). In terms of the dimensionless coordinate,
time, and wave amplitude introduced in Ref. \cite{UJP_2014}, the
coefficients $a_{n}\ldots$ are all real and are functions of one
dimensionless depth parameter $kh$, $k$ being the carrier wave
number and $h$ being the undisturbed fluid depth. The corresponding
explicit formulas for these coefficients are given in Ref.~\cite{PRL_2015}.

\begin{figure}[t]
\includegraphics[width=\columnwidth]{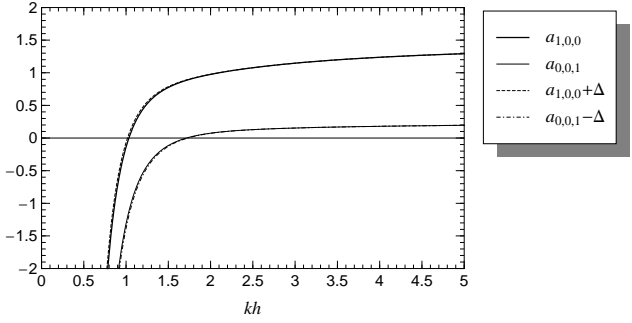}
\caption{\label{fig:delta}The effect of correction $\Delta$ on the coefficients $a_{1,\,0,\,0}$ and $a_{0,\,0,\,1}$.
(Reproduction of Fig.~15 from Ref.~\cite{UJP_2014}.)}
\end{figure}

Equation (\ref{eq:Slunyaev_full}) was derived by Slunyaev
\cite{Slunyaev_2005} (2005). It takes into account nonlinear and
nonlinear-dispersive terms in the next order of smallness as
compared to Eq.~(\ref{eq:Sedletsky}). Note that here we use
different notation for the variables and coefficients as compared to
the original notation used by Slunyaev in Ref.~\cite{Slunyaev_2005}.
In deriving his more general HONLSE model, Slunyaev also introduced
a correction to the coefficients $a_{1,\,0,\,0}$ and $a_{0,\,0,\,1}$
derived earlier in Ref.~\cite{SedletskyJETP2003}:
\begin{equation}\label{eq:Delta_pm}
\begin{split}
\widetilde{a}_{1,\,0,\,0}&=a_{1,\,0,\,0}+\Delta,\\
\widetilde{a}_{0,\,0,\,1}&=a_{0,\,0,\,1}-\Delta,
\end{split}
\end{equation}
where the correction $\Delta$ is expressed as follows
\cite{UJP_2014}
\begin{align}
\Delta =&\;-\frac{1}{32\sigma^3\nu}\Bigl(\bigl(\sigma^2-1\bigr)^4\bigl(3\sigma^2+1\bigr)k^3h^3\nonumber \\
 &\;-\sigma\bigl(\sigma^2-1\bigr)^2\bigl(5\sigma^4-18\sigma^2-3\bigr)k^2h^2 \\
 &\;+ \sigma^2\bigl(\sigma^2-1\bigr)^2\bigl(\sigma^2-9\bigr)kh + \sigma^3\bigl(\sigma^2-1\bigr)\bigl(\sigma^2-5\bigr)\Bigr).\nonumber
\end{align}
Here $\sigma=\tanh(kh)$ and
\begin{equation}
\nu =\left(\sigma^2-1\right)^2 k^2 h^2 - 2\sigma\left(\sigma^2+1\right)kh + \sigma^2.
\end{equation}
It was demonstrated in our earlier work \cite{UJP_2014} that the
correction $\Delta$ is small as compared to the values of
coefficients $a_{1,\,0,\,0}$ and $a_{0,\,0,\,1}$. It is well seen
from Fig.~\ref{fig:delta} reproduced here for clarity. Actually,
this was the reason why this correction was originally ignored by
Sedletsky in Ref.~\cite{SedletskyJETP2003}. Since this correction is
small, we deliberately ignore it in Ref.~\cite{PRL_2015} as well.

Slunyaev's model (\ref{eq:Slunyaev_full}) has one major drawback:
the coefficients $\widetilde{a}_{2,\,0,\,0}$,
$\widetilde{a}_{0,\,0,\,2}$, and $\widetilde{a}_{1,\,0,\,1}$ are
asymptotically divergent at $kh\rightarrow\infty$. To avoid this
problem in the vicinity of $kh=1.363$, where the third-order
nonlinear coefficient $a_{0,\,0,\,0}$ vanishes (modulation
instability threshold), Slunyaev renormalized high-order terms to
get a simplified HONLSE model (\ref{eq:Slunyaev_truncated}), which
we will address to as truncated Slunyaev's equation. It is important
to note that this truncated model was formulated by Slunyaev only
for the small vicinity of $kh=1.363$. Grimshaw and Annenkov
\cite{Grimshaw_2011} used truncated Slunyaev's equation
(\ref{eq:Slunyaev_truncated}) to derive a general solitary wave
solution that transforms to a bright soliton in the limiting case.
They proved that this bright soliton exists in the small vicinity of
$kh \approx 1.363$.

Note that Eq.~(\ref{eq:Sedletsky}) is essentially different from
Eq.~(\ref{eq:Slunyaev_truncated}): it takes into account the
third-order dispersion (term with $u_{\chi\chi\chi}$) but neglects
the fifth-order nonlinearity (term with $|u|^4 u$). In Ref.~\cite{PRL_2015}, we start from Eq.~(\ref{eq:Sedletsky})
to derive a new bright soliton solution that exists below the
modulation instability threshold. This bright soliton solution has
the following form:
\begin{align}
u_B(\chi,\,\tau)\!=&\,u_0\sech\bigl(K(\chi-\chi_0-V\tau)\bigr)\;\E^{\I\kappa\chi-\I\Omega\tau},\label{eq:soliton_bright}
\end{align}
where $\chi_0$ is the soliton's arbitrary initial position and
\begin{subequations}\label{eq:BS_HONLSE}
\begin{gather}
K = |u_0|\sqrt{S},\quad S = -\frac{a_{1,\,0,\,0}+a_{0,\,0,\,1}}{6a_{3}},\label{eq:K_HONLSE}\\
\kappa = \frac{a_{1,\,0,\,0}+a_{0,\,0,\,1}-6a_{3}a_{0,\,0,\,0}}{12a_{3}a_{0,\,0,\,1}},\label{eq:kappa_HONLSE}\\
\Omega = \kappa a_1+\frac{1}{2}\Bigl(K^2\left(1-6\kappa a_{3}\right)-\kappa^2\left(1-2\kappa a_{3}\right)\Bigr),\\
V = a_1 -\kappa + \left(3\kappa^2 - K^2\right) a_{3}.
\end{gather}
\end{subequations}
In this case formula (\ref{eq:soliton_bright}) represents a
one-parametric family of solutions with variable soliton amplitude
$u_0$. It is well seen from Eq.~(\ref{eq:K_HONLSE}) that the
presence of the third-order dispersion in the HONLSE model is
absolutely important for the existence of solution
(\ref{eq:soliton_bright}). This solution was shown to exist in the
following depth range (condition $S\geqslant0$) \cite{PRL_2015}:
\begin{equation}\label{eq:S_range}
0.763\lesssim kh\lesssim1.222.
\end{equation}
The condition of narrow spectrum $\kappa\ll 1$ (slow modulation)
imposes an additional restriction on the allowable depth values,
which should lie in a narrow range around $kh\approx 1.249$, where
$\kappa=0$. In the numerical calculations presented in Ref.~\cite{PRL_2015}, we selected the depth parameter equal to
$kh=1.2$, at which we have $\kappa\approx 0.230$.

The purpose of this short note is to demonstrate that the new bright
soliton solution (\ref{eq:soliton_bright}) presented in Ref.~\cite{PRL_2015} is not significantly modified by
Slunyaev's correction $\Delta$ and that it also exists in the
framework of a more general HONLSE model with the fifth-order
nonlinearity taken into account.

{\it The effect of Slunyaev's correction.} First, it is important to
note that the correction $\Delta$ does not modify the parameters $S$
and $K$ in formula (\ref{eq:K_HONLSE}) and the range of solution
existence (\ref{eq:S_range}), since it is canceled out in the sum
$a_{1,\,0,\,0}+a_{0,\,0,\,1}$, inasmuch as $\Delta$ makes equal
corrections with opposite signs to the both coefficients, as seen
from Eq.~(\ref{eq:Delta_pm}). Therefore, the correction $\Delta$ has
effect only on the value of parameter $\kappa$, where it appears in
the denominator of expression (\ref{eq:kappa_HONLSE}) through the
coefficient $a_{0,\,0,\,1}$. For $kh=1.2$, the corrected value of
parameter $\kappa$ is $0.216$, which makes only a 6\% difference to
our original estimate $\kappa\approx 0.230$. Thus, neglecting
Slunyaev's correction $\Delta$ is quite legitimate within the limits
of accuracy of HONLSE model (\ref{eq:Sedletsky}), but it should be
taken into account in the next order of smallness introduced in a
more general HONLSE model (\ref{eq:Slunyaev_full}).

{\it The effect of the fifth-order nonlinearity.} The bright soliton
solution in form (\ref{eq:soliton_bright}) and (\ref{eq:BS_HONLSE})
does not exist in the framework of truncated Slunyaev's equation
(\ref{eq:Slunyaev_truncated}) because the condition $a_3\ne 0$ does
not hold true in that case. Therefore, we need to consider a more
general equation with the $a_3$ term preserved:
\begin{multline}\label{eq:HONLSE5}
u_\tau = - a_1 u_\chi - \I a_2 u_{\chi\chi}+\I a_{0,\,0,\,0}|u|^2 u \\
+ \Bigl(a_3 u_{\chi\chi\chi} - a_{1,\,0,\,0}u_{\chi}|u|^{2} - a_{0,\,0,\,1}u^{2}u^{*}_{\chi}\Bigr)\\
+\I a_{0,\,0,\,0,\,0,\,0}|u|^4 u.
\end{multline}
In terms of dimensionless variables used in Ref.~\cite{PRL_2015}, the fifth-order coefficient is expressed as
\begin{equation}
a_{0,\,0,\,0,\,0,\,0} \equiv -\frac{\alpha_{31}}{\omega k^4 c}.
\end{equation}
The expression for the coefficient $\alpha_{31}$ is given in
Slunyaev's work \cite{Slunyaev_2005}, and the expression for the
dimensionless phase speed $c$ is given in Ref.~\cite{PRL_2015}.
Here, the sign minus at the coefficient $\alpha_{31}$ is due to the
fact that Slunyaev used the conjugate carrier wave basis as compared
to our notation. For $kh = 1.2$, we have $\alpha_{31}\approx 0.30$.

\begin{figure}[t]
\includegraphics[width=\columnwidth]{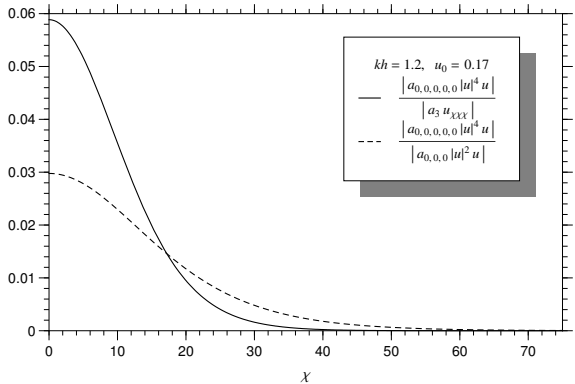}
\caption{\label{fig:T5toT3}Ratios of the fifth-order nonlinear term to the third-order dispersive term and the third-order nonlinear term
for the wave envelope in the form of bright soliton (\ref{eq:soliton_bright}) with parameters (\ref{eq:BS_HONLSE}) and $u_0=0.17$ at $kh=1.2$.}
\end{figure}

\begin{figure}[t]
\includegraphics[width=\columnwidth]{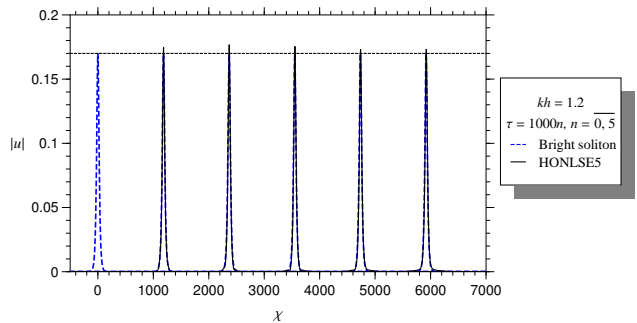}
\caption{\label{fig:HONLSE5_kh1.2} The effect of the fifth-order nonlinearity on the evolution of the bright HONLSE soliton with $u_0 = 0.17$ ($kh=1.2$)
calculated by numerical integration of Eq.~(\ref{eq:HONLSE5}) using the split-step Fourier technique.}
\end{figure}

To analyze the effect of the fifth-order nonlinear term on the
evolution of bright HONLSE soliton (\ref{eq:soliton_bright}), we
used this wave form as the initial condition in
Eq.~(\ref{eq:HONLSE5}) at $kh=1.2$. Figure \ref{fig:T5toT3}
demonstrates that in this case the fifth-order term makes only a
small perturbation to the third-order dispersive term and the
third-order nonlinear term. It can also be seen that the inputs of
the third-order dispersive and nonlinear terms have the same order
of magnitude. This fact proves once again the the third-order
dispersion cannot be neglected in the correct description of wave
evolution at $kh=1.2$. Figure~\ref{fig:HONLSE5_kh1.2} shows the
evolution of the initial envelope in the form of bright soliton
computed using the split-step Fourier technique described in Ref.
\cite{UJP_2014}. The initial values of parameters were selected to
be exactly the same as in Ref.~\cite{PRL_2015}. It can
be seen that the numerical solution (HONLSE5) exhibits slight
oscillations around the initial amplitude $u_0=0.17$. This behavior
was earlier addressed by us as a quasi-soliton behavior
\cite{UJP_2014}. Here we observe exactly the same effect as in the
case of NLSE solitons used as initial conditions in a more general
HONLSE model (\ref{eq:Sedletsky}) with high-order terms taken into
consideration. Thus, the bright soliton solution
(\ref{eq:soliton_bright}) is preserved in a more general HONLSE
model (\ref{eq:HONLSE5}) in the form of quasi-soliton with slowly
varying amplitude. On short time intervals, it behaves as the true
soliton, in view of the smallness of the fifth-order contribution.


{\it Modulation instability threshold.} In the framework of
conventional NLSE equation (when the second row of
Eq.~(\ref{eq:Sedletsky}) is neglected), the modulation instability
threshold lies at the point $kh = 1.363$, where $a_{0,\,0,\,0} = 0$.
In the framework of a more general HONLSE model
(\ref{eq:Sedletsky}), the modulation instability threshold slightly
shifts to higher $kh$, depending on the amplitude $u_0$ of a
homogeneous solution. In particular, at $u_0=0.17$, the modulation
instability threshold shifts to $kh\approx 1.365$ \cite{PRL_2015}.
In the framework of HONLSE model (\ref{eq:HONLSE5}) with the
fifth-order nonlinear term, the modulation instability criterion
should be modified appropriately. A homogenous solution to
Eq.~(\ref{eq:HONLSE5}) has the following form:
\begin{equation}
u(\chi,\,\tau)=u_0\E^{\I\left(a_{0,\,0,\,0}|u_0|^2+a_{0,\,0,\,0,\,0,\,0}|u_0|^4\right)\tau}.
\end{equation}
Its instability criterion can be determined by introducing a small
perturbation to the amplitude $u_0$:
\begin{gather}
u(\chi,\,\tau)=\bigl(u_0+\epsilon(\chi,\,\tau)\bigr)\E^{\I \left(a_{0,\,0,\,0}|u_0|^2+a_{0,\,0,\,0,\,0,\,0}|u_0|^4\right)\tau},\nonumber\\
\epsilon(\chi,\,\tau)=\epsilon_0^{+}\E^{\I\kappa\chi-\I\Omega\tau}+\epsilon_0^{-}\E^{-\I\kappa\chi+\I\Omega^*\tau}.
\end{gather}
Here, we assume the perturbation frequency $\Omega$ to be
complex-valued and the perturbation wave number $\kappa$ to be real.
Substituting this ansatz in Eq.~(\ref{eq:HONLSE5}) leads to the
following dispersion relation between $\Omega$  and $\kappa$:
\begin{gather}\label{eq:MI_Omega}
\Omega = \left(a_1+ a_{1,\,0,\,0}|u_0|^2\right)\kappa + a_3 \kappa^3 \pm \kappa\sqrt{R},\\
R = 2a_2 \left(a_{0,\,0,\,0}+2a_{0,\,0,\,0,\,0,\,0}|u_0|^2\right)|u_0|^2 \nonumber\\
+\, a_{0,\,0,\,1}^2|u_0|^4 + a_2^2\kappa^2\nonumber.
\end{gather}
A homogeneous solution is modulationally unstable when the
perturbation exponentially grows with time. This happens when
$\Im\Omega>0$, which effectively requires the radicand $R$ in Eq.
(\ref{eq:MI_Omega}) to be negative. This condition is satisfied when
\begin{equation}\label{eq:MI}
a_{0,\,0,\,0} < - \left(a_{0,\,0,\,1}^2+2a_{0,\,0,\,0,\,0,\,0}\right) |u_0|^2,
\end{equation}
where we took into account that $a_2 = \frac{1}{2}$. Criterion
(\ref{eq:MI}) was derived earlier in Ref. \cite{Slunyaev_2005}. The
fifth-order nonlinearity makes a significant input in the modulation
instability criterion, so that the instability threshold shifts in
the direction of smaller $kh$, inasmuch as the expression
$a_{0,\,0,\,1}^2+2a_{0,\,0,\,0,\,0,\,0}$ is negative at
$kh\gtrsim1.20$. In particular, at $u_0=0.17$, the modulation
instability threshold shifts to $kh\approx 1.353$. This point,
however, lies far above the region of existence of the bright
soliton solution discussed above, and all our predictions regarding
the evolution of this new solution remain valid.

{\it Conclusions.} In this short note we proved that
\begin{enumerate}
\item Slunyaev's correction to Sedletsky's coefficients is small and
can readily be ignored within the limits of accuracy of HONLSE model
(\ref{eq:Sedletsky}).

\item At $kh=1.2$, the third-order dispersion and the third-order nonlinearity
make the comparative contributions to the evolution equation and,
therefore, the third-order dispersion cannot be ignored.

\item At $kh=1.2$, the effect of the fifth-order nonlinearity is small as
compared to the third-order dispersion and the third-order
nonlinearity. It manifests in the slight oscillations of the soliton
amplitude around the undisturbed initial level (the so-called
quasi-soliton behavior).

\item The fifth-order nonlinearity makes a significant correction to
the modulation instability threshold, so that it shifts in the
direction of smaller $kh$. However, this effect is important only in
the small vicinity of $kh \approx 1.363$ and is insignificant at
smaller depths, in particular, at $kh = 1.2$.

\item The criterion of existence of HONLSE bright solitons and the criterion of modulation instability are two different and independent criteria, and this conclusion holds true in the case of the fifth-order nonlinearity.
\end{enumerate}

\end{document}